\documentclass{rnaastex}

\begin{document}

\title{An improved quantification of HD 147379 b}

\correspondingauthor{Fabo Feng}
\email{fengfabo@gmail.com,f.feng@herts.ac.uk}

\author[0000-0001-6039-0555]{Fabo Feng}
\altaffiliation{Centre for Astrophysics Research, University of Hertfordshire, College Lane, AL10 9AB, Hatfield, UK}

\author{Hugh R. A. Jones}
\affiliation{Centre for Astrophysics Research, University of Hertfordshire, College Lane, AL10 9AB, Hatfield, UK}

\author{Mikko Tuomi}
\affiliation{Centre for Astrophysics Research, University of Hertfordshire, College Lane, AL10 9AB, Hatfield, UK}

%% Note that RNAAS manuscripts DO NOT have abstracts.
\keywords{methods: statistical -- methods: data analysis -- techniques: radial velocities -- stars:  individual: HD147379}

\section{}
%\vspace{-2in}
%\section{Data}
%\section{Result}
A Neptune-like planet is reported to orbit around HD~147379 (GJ617A) in its conservative habitable zone with a period of 86.54\,d and a minimum mass of 25\,$M_\oplus$ based on an analysis of the CARMENES radial velocity (RV) data \citep{reiners17} and supported by HIRES/Keck data \citep{butler17}. Signals with periods of about 21.1\,d and 10.6\,d are also found to be significant, although are interpreted as activity-induced variation. \cite{pepper18} confirm this analysis with KELT photometric data where a 22d periodic signal is identified as the rotational period of HD~147379.

\begin{figure}[b]
  \centering
\includegraphics[scale=0.5]{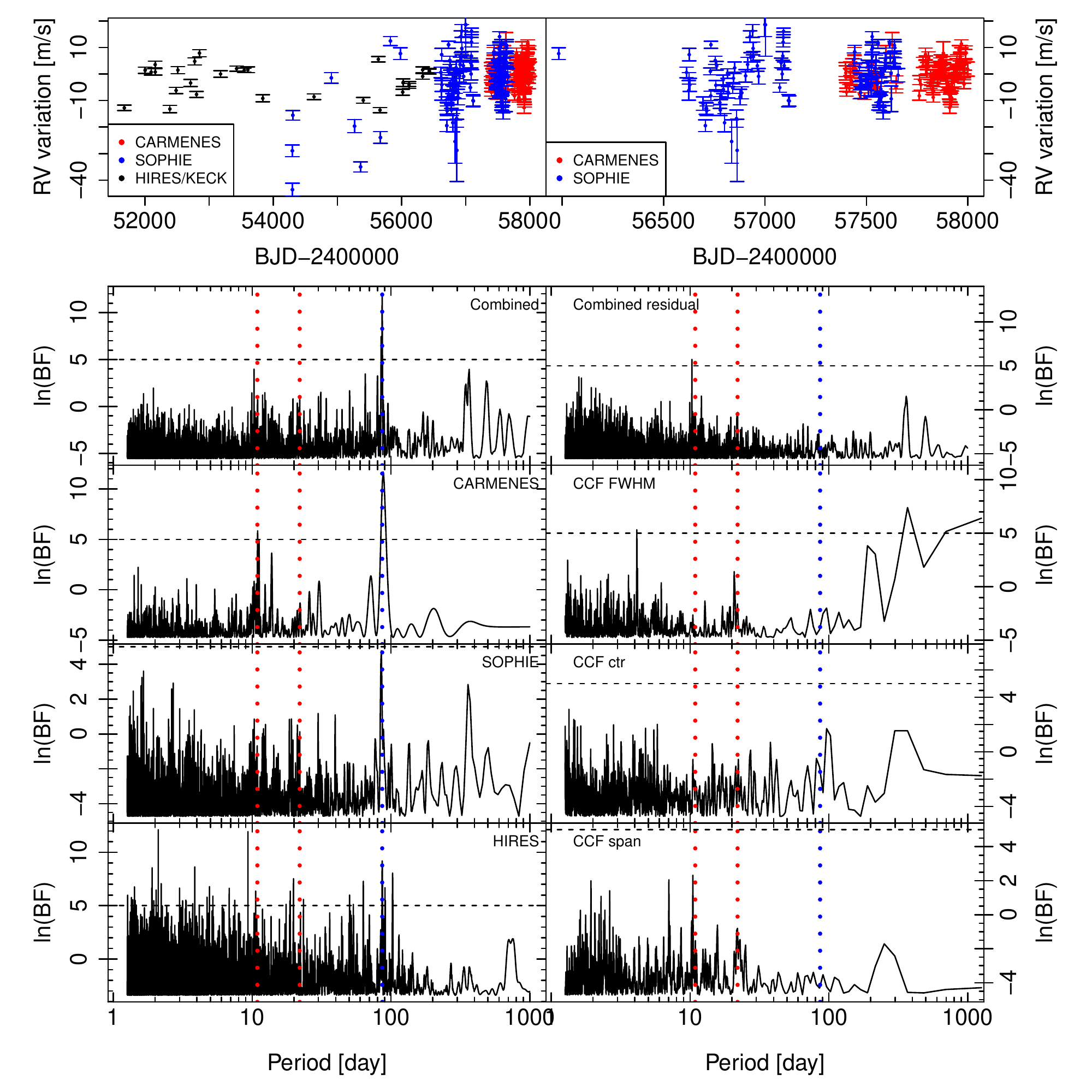}
\caption{RV data sets from CARMENES, SOPHIE, and HIRES (top panels) and the BFPs for RVs, RV residuals, and some activity indicators. The red dotted lines show the rotation period of 22\,d and its harmonic around 11\,d. The blue dotted lines show the signal at a period of 86.5\,d. The threshold of ln(BF)$=5$ is shown by the dashed horizontal lines.}
\label{fig1}
\end{figure}

To further investigate the nature of these signals and constrain the signals better, we reanalyze the CARMENES and HIRES/KECK data, and have also analyzed the SOPHIE data \citep{perruchot08} using the Agatha software \citep{feng17a} in combination with MCMC posterior sampling \citep{feng16}. Considering that RV data is typically contaminated by red noise \citep{feng16}, we use the red noise periodogram within Agatha called ``Bayes factor periodogram'' (BFP) to identify periodic signals. In Fig. \ref{fig1} we show the BFPs for the RV data of CARMENES, HIRES, SOPHIE, the combined data, and some of the activity indicators of the CARMENES data set. In all RV data sets, we find significant power at 86.5\,d, confirming the Keplerian nature of this signal. This signal does not appear in activity indices. The rotation period at 22\,d and its harmonic is strong in RVs and indices, consistent with the rotation period found in photometric data. 

We model the noise in CARMENES and SOPHIE data using the first moving average model, and model the noise in the HIRES data using the white noise model according to the Bayesian model comparison \citep{feng16}. By fitting one planet model to the combined RV data sets, we derive an orbital period\footnote{The confidence level is estimated by the 10\% and 90\% quantiles of the posterior distribution. The optimal value is estimated at the maximum {\it a posteriori} (MAP). } of $P=86.53^{+0.09}_{-0.09}$\,d, minimum mass of $m\sin{i}=27.07_{-6.8}^{+3.42}$\,$M_\oplus$, semi amplitude of $K=5.63^{+0.27}_{-1.24}$\,m/s, semi-major axis of $a=0.319^{+0.019}_{-0.019}$\,au, eccentricity of $e=0.02^{+0.03}_{-0.01}$, and angular parameters ranging from 0 to 2$\pi$. We also find the jitter (or excess white noise) for CARMENES, SOPHIE, and HIRES, are 0.319$^{+0.019}_{-0.019}$\,m/s, 7.88$^{+1.99}_{-0.91}$\,m/s, and 3.86$^{+2.04}_{-0.95}$\,m/s, respectively. The red noise in the CARMENES data has an amplitude of $0.83^{+0.16}_{-0.44}$\,m/s and a logarithmic time scale (in unit of day) of 1.77$^{+2.01}_{-0.47}$. The red noise in the SOPHIE data has an amplitude of 0.44$^{+0.27}_{-0.03}$\,m/s and a logarithmic time scale of 7.90$^{+0.03}_{-4.63}$. The jitter of CARMENES data determined in this work is significantly lower than the value of 3.7$^{+0.4}_{-0.2}$\,m/s determined by \cite{reiners17}, suggesting that the noise in CARMENES data is dominated by time-correlated noise. The total noise-induced RV variations are about 3.1, 8.3, and 3.9\,m/s for the CARMENES, SOPHIE, and HIRES data, respectively. This means that the uncertainty in the SOPHIE data is significantly underestimated, compared with the CARMENES and HIRES data and that the precision of CARMENES is similar to that of HIRES/KECK. 

The estimated uncertainty of the parameters is larger than that estimated by \cite{reiners17}  probably due to our adoption of a higher confidence level and the inclusion of red noise components in the model. On the other hand, the uncertainty of semi-major axis determined by \cite{reiners17} is probably underestimated due to neglecting stellar mass uncertainty. 

In summary, our analysis of three independent data sets confirms the existence of HD~147379~b and improves the parameter estimation through modeling of red noise. The multiple RV data sets for HD147379 provide a benchmark case for RV data analyses. Based on our analysis of the HD147379 data, CARMENES RVs are dominated by red noise and thus need to incorporate a red noise model.

\acknowledgments
We thank Mathias Zechmeister for sharing the CARMENES data. 

\bibliographystyle{aasjournal}
\bibliography{nm}
% \end{CJK*}
\end{document}